\documentclass{aa}
\usepackage[dvips]{graphicx}

\begin{document}
 
\title{The periods of the intermediate polar RX J0153.3+7446}
 
\titlerunning{The intermediate polar RX J0153.3+7446}
 
\author{A.J. Norton \and J.D. Tanner}
 
\authorrunning{Norton \& Tanner}
 
\offprints{A.J. Norton, a.j.norton@open.ac.uk}
 
\institute{Department of Physics and Astronomy, The Open University,
        Walton Hall, Milton Keynes MK7 6AA, U.K.}
 
\date{Accepted ???
      Received ???}
 
\abstract{We present the first optical photometry of the counterpart to the 
candidate intermediate polar RX J0153.3+7446. This reveals an optical pulse period
of $2333\pm5$~s. Reanalysis of the previously published {\em ROSAT} X-ray data
reveals that the true X-ray pulse period is probably $1974\pm30$~s, rather than
the 1414~s previously reported. Given that the previously noted orbital period 
of the system is 3.94~h, we are able to 
identify the X-ray pulse period with the white dwarf spin period and the optical
pulse period with the rotation period of the white dwarf in the binary reference
frame, as commonly seen in other intermediate polars. We thus confirm that 
RX J0153.3+7446 is indeed a typical intermediate polar.

}

\maketitle

\section{Introduction}

Intermediate polars are cataclysmic variables, characterised by pulsed X-ray emission 
which reflects the rotation period of an accreting magnetic white dwarf. Over 
two dozen confirmed systems are currently known (for a recent list see Norton, 
Wynn \& Somerscales 2004) and of these, around a quarter were discovered by 
{\em ROSAT} during its all sky survey. In an important paper, Haberl \& Motch (1995)
reported the first {\em ROSAT} observations of six intermediate polar candidates. Five 
of these systems have gone on to be well studied. RE0751+14 (PQ Gem; see Duck et al
1994, Potter et al 1997), RX J0028.8+5917 (V709 Cas; see Norton et al 1999, de Martino 
et al 2001) and RX J0558.0+5353 (V405 Aur; see Allan et al 1996, Evans \& Hellier 2004) 
are now confirmed as typical intermediate polars with a wealth of published 
observations; RX J1712.6--2414 (V2400 Oph;  see Buckley et al 1997; Hellier \& Beardmore
2002) is the first confirmed stream-fed intermediate polar; and RX J1914.4+2456 
(V407 Vul; see Cropper et al 1998, Ramsay et al 2002, 2005, Norton, Haswell \& Wynn
2004) has excited much controversy as a possible ultra-compact binary, although its nature
is still in doubt. 

However, the sixth object reported by Haberl \& Motch (1995), namely RX J0153.3+7446, 
has been virtually ignored ever since. They showed an X-ray pulse profile of the 
system, claiming the pulse period as 1414s, and mentioned that a publication on
optical observations of the object was in preparation. Such a publication has never 
appeared, although the on-line cataclysmic variable catalogue (Downes et al 2001)
lists an optical counterpart in Cassiopeia with V=16.4 at RA 01:53:20.9, Dec +74:46:22
and mentions an orbital period of 0.16415~d credited to a private communication from 
John Thorstensen. The only published observation of RX J0153.3+7446 is an 
identification spectrum by Liu \& Hu (2000) which confirms its cataclysmic variable 
nature by virtue of the typical emission line spectrum.

Optical photometry of intermediate polars usually shows modulation at the 
X-ray pulse period (i.e. the spin period of the white dwarf in most cases)
and/or the beat period (i.e. the spin period of the white dwarf in the binary
reference frame). The latter modulation, where seen, is presumed to be due to
reprocessed X-ray emission originating from the illuminated surface of the 
donor star or some other structure fixed in the binary reference frame. An
orbital photometric modulation is often seen as well, presumably depending 
on the inclination angle of the system.

\section{Observations}

In an attempt to confirm the nature of RX J0153.3+7446, we have carried out
optical photometric observations of its 16th magnitude counterpart. The 
unfiltered photometry was carried out on the night of 2005 November 24th
using the 40cm Alan Cooper Telescope, a Meade LX200 Schmidt Cassegrain
telescope at the Open University's George Abell Observatory in Milton
Keynes, U.K. We obtained over 350 exposures, each of 30s, between 
19:08UT and 22:54UT. There is a deadtime of $\sim 5$s between each
exposure whilst the full frame was read out. Images were acquired using 
an SBIG STL1001E CCD camera, giving an image scale of 1.24 arcseconds per 
pixel. The counterpart to RX J0153.3+7446 is clearly detected with a 
signal-to-noise ratio of typically $\sim 20$ on each 
frame, separated from a nearby 11th magnitude star (TYC 4322-169-1) by 
around 12 arc seconds (see Figure 1). The seeing was very stable at
$\sim 2.5$ arc seconds throughout the observations.

Using the MaximDL (v4) software, all images were bias and dark current 
subtracted, then flat fielded using dome flats. Aperture photometry was 
performed with respect to the nearby star TYC 4322-255-1 which has V=12.058, 
B=13.005 according to {\em Vizier}. Comparison with a check star (for which 
also see Figure 1) confirmed that the magnitude of the reference star 
was constant for the duration of the observations. 

\begin{figure}[t]
\begin{center}
\includegraphics[scale=0.4]{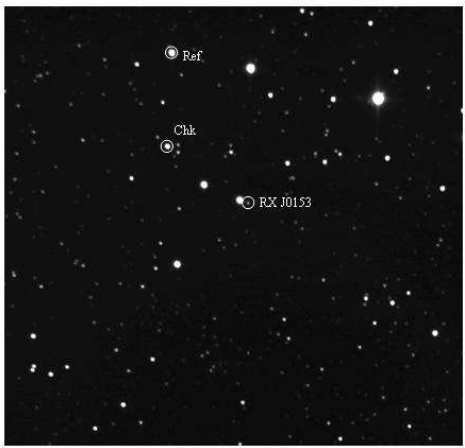}
\caption{The field of RX J0153.3+7446. The reference star and check star used
for the photometry are indicated. The field shown here is about 12 arcminutes 
across, North is up and East to the left.}
\end{center}
\end{figure}

\section{Results}

The optical lightcurve of RX J0153.3+7446, spanning almost 4 hours, is shown
in Figure 2. A long term modulation is clearly present, consistent with 
the 3.94 h orbital period listed in the Downes et al (2001) catalogue.
Superimposed on top of this are around 6 cycles of a further modulation, 
with an amplitude of $\sim 0.1$ magnitudes. A Fourier transform of the 
lightcurve reveals a strong peak at $4.286 \times 10^{-4}$~Hz, see 
Figure 3. The period of this additional modulation is therefore 2333s, 
with an uncertainty of $\sim 5$s. The lightcurve folded at this 
period, and binned into 25 phase bins, is shown in Figure 4. Error bars 
represent the standard error on the mean within each bin, averaged over 
typically 15 observations each.

\begin{figure}[t]
\begin{center}
\includegraphics[scale=0.3,angle=-90]{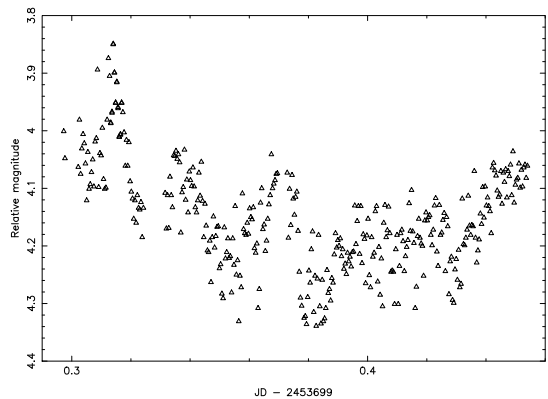}
\caption{The unfiltered optical lightcurve of RX J0153.3+7446. The magnitudes 
are relative to that of TYC 4322-255-1 which has V=12.058, B=13.005 according 
to {\em Vizier}.}
\end{center}
\end{figure}

\begin{figure}[t]
\begin{center}
\includegraphics[scale=0.3,angle=-90]{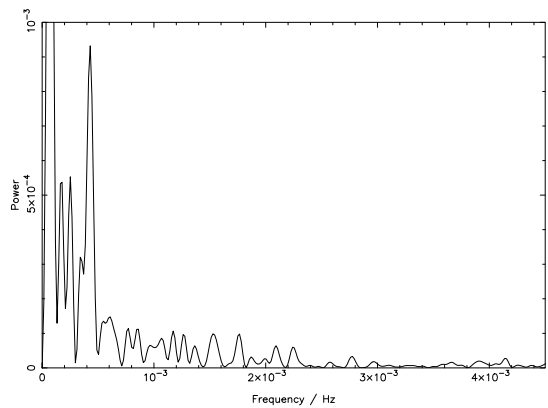}
\caption{The power spectrum of the RX J0153.3+7446 optical lightcurve. Significant 
power is seen at a frequency of $4.286\times10^{-4}$~Hz, corresponding to a 
period of 2333s.}
\end{center}
\end{figure}

\begin{figure}[t]
\begin{center}
\includegraphics[scale=0.3,angle=-90]{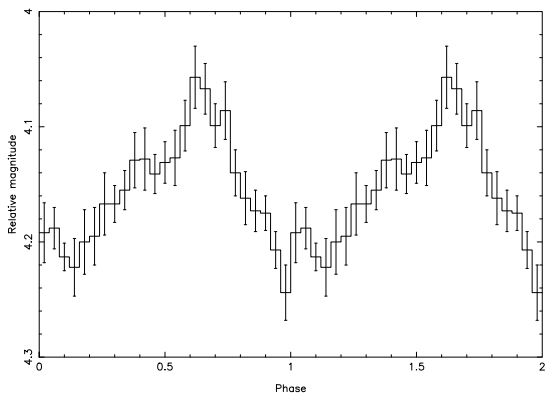}
\caption{The optical lightcurve of RX J0153.3+7446, folded at 2333s and 
binned into 25 phase bins. Magnitudes are relative to  TYC 4322-255-1 and
the phasing is arbitrary.}
\end{center}
\end{figure}

There is no sign of a modulation at the previously reported X-ray pulse
period of 1414s, and nor is the period we see a harmonic or sideband of 
that and the reported orbital period.

Although these are the first photometric observations to be reported from
the George Abell Observatory, following its recent refurbishment, the performance
of the telescope plus CCD system was verified a few days prior to these 
observations when we observed another of the intermediate polars from Haberl \& 
Motch's original paper. On 2005 November 19th we obtained $\sim 2.5$h of 
unfiltered photometry of the 14th magnitude intermediate polar V405 Aur. The 
lightcurve of those observations clearly shows the well established 
(Allan et al 1996) double peaked pulse profile with a 545s pulse period and 
peak-to-peak amplitude of $\sim 0.1$ magnitude. Hence we are confident 
in the stability and reliability of our observing system.

\section{Re-analysis of the {\em ROSAT} observations}

In an attempt to shed light on the discrepancy between our optical
photometric period and the X-ray period reported by Haberl \& Motch (1995), we
have re-analysed the original {\em ROSAT} observations of RX J0153.3+7446. 
These comprise two observations, each of $\sim 8000$s on-source exposure, 
obtained in 1991 and 1992. As noted by Haberl \& Motch, the first observation 
is not suitable for period determination as it comprises many short segments, 
spread over a year. The lightcurve of the observation from 1992 March is
more suitable for period determination, but even so it comprises a total of 
only 186 photons from RX J0153+7446, distributed over six satellite orbits and 
spanning $\sim$58~ksec. 

\begin{figure}[t]
\begin{center}
\includegraphics[scale=0.3,angle=-90]{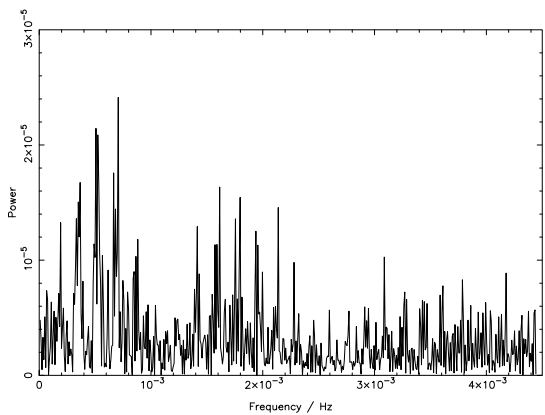}
\caption{The power spectrum of the RX J0153.3+7446 {\em ROSAT} X-ray lightcurve. 
Significant power is seen at frequencies of $7.05 \times 10^{-4}$~Hz (i.e. 1414s)
and at  $5.06 \times 10^{-4}$~Hz, corresponding to a period of 1974s.}
\end{center}
\end{figure}

\begin{figure}[t]
\begin{center}
\includegraphics[scale=0.3,angle=-90]{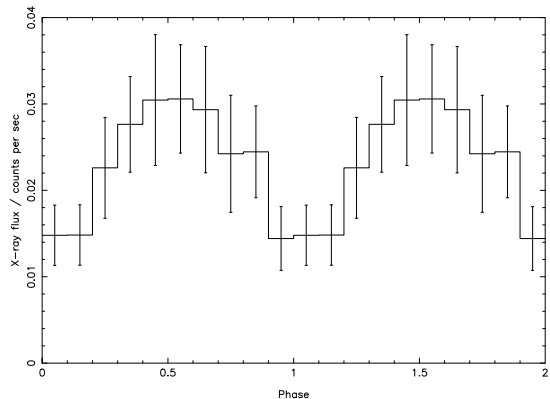}
\caption{The X-ray lightcurve of RX J0153.3+7446, folded at 1974s and 
binned into 10 phase bins. The phasing is arbitrary.}
\end{center}
\end{figure}

The power spectrum of this lightcurve is shown in Figure 5. Although the highest 
peak in the power spectrum (at $7.05 \times 10^{-4}$~Hz) does indeed correspond 
to a period of $\sim 1414$s, as reported by Haberl \& Motch, the power spectrum 
is understandably noisy. Moreover, the second highest peak (at $5.06 \times 
10^{-4}$~Hz) is clearly an alias of the 1414s period, and only marginally less
preferred. This corresponds to a period of $1974 \pm 30$s. The X-ray lightcurve
folded at this period is shown in Figure 6. If we assume that 1974~s
is the true X-ray pulse period of RX J0153+7446, then the optical photometric
period which we have found is harmonically related to this and the proposed
orbital period via:

\begin{eqnarray}
\frac{1}{P_{\rm spin}} - \frac{1}{P_{\rm orb}} & = & \frac{1}{P_{\rm beat}} \\ 
\frac{1}{(1974 \pm 30){\rm s}} - \frac{1}{3.94{\rm h}} & \sim & \frac{1}{(2333 \pm 5){\rm s}} \nonumber
\end{eqnarray}

This is the usual way in which the white dwarf spin period, orbital period
and beat period are related in intermediate polars, and reflects the fact
that the dominant optical photometric modulation is due to reprocessing of 
the X-ray pulse from a site fixed in the binary reference frame, such as 
the face of the companion star.

\section{Conclusions}

We have detected a strong optical photometric modulation from RX J0153+7446 at a 
period of $2333\pm5$s. Having reanalysed the {\em ROSAT} X-ray observation of 
the source, we find that the true X-ray pulse period is probably $1974\pm30$sec.
Assuming that the X-ray period represents the white dwarf spin period and the optical 
period represents the rotation period of the white in the binary reference frame,
then the two are in agreement with the reported orbital period for this system
of 3.94h. RX J0153+7446 is thus confirmed as an intermediate polar.

The spin to orbital period ratio of RX J0153.3+7446 is therefore 
0.14, which is one of the highest amongst intermediate polars above the 
period gap. From the results presented in Norton, Wynn \& Somerscales (2004)
we can estimate the magnetic moment of the white dwarf in this systems as 
$\sim 6 \times 10^{33}$~G~cm$^{3}$. This places RX J0153.3+7446 at the high
end of the intermediate polar magnetic moment distribution and suggests
it may be a candidate for exhibiting polarized emission and also episodes
of stream-fed accretion. Further X-ray observations are encouraged to confirm and 
refine the X-ray pulse period and orbital period of this system.

\begin{acknowledgements}
The George Abell Observatory has been funded by grants from the Open University
Capital Equipment Fund and General Purposes Fund. The research reported here 
has made use of the SIMBAD database, operated at CDS, Strasbourg, France.
\end{acknowledgements}

\end{document}